\documentclass{elsart}
\usepackage{psfig}
\usepackage{natbib}
\begin{document}

\runauthor{Edelson}

\begin{frontmatter}

\title{X-Ray Variability of NLS1s and BLS1s}

\author[UCLA]{Rick Edelson}

\address[UCLA]{UCLA Department of Physics \& Astronomy; Los Angeles, CA 
90095-1562; USA}

\begin{abstract}

Evenly-sampled hard X-ray monitoring was obtained with RXTE for one NLS1 (Akn 
564) and four BLS1s.
The variability amplitude of the NLS1 was no larger than the mean of the BLS1s, 
and the NLS1 showed stronger variability in the harder portion of the RXTE band, 
while the BLS1s were more strongly variable in the relatively soft part of the 
band.
This contribution discusses possible explanations for these surprising results, 
including possible calibration errors or systematic differences between long and 
short time scale variability in NLS1s and BLS1s.

\end{abstract}

\begin{keyword}
galaxies: active; quasars: general; quasars: absorption lines; X-rays: galaxies
\end{keyword}

\end{frontmatter}


\section{Introduction}

Narrow-Line Seyfert 1s (NLS1s; see, e.g., Boller, Brandt \& Fink 1996) 
show much steeper and variable soft X-ray emission and narrower optical permitted lines than Broad-Line Seyfert 1s (BLS1s).
This is in many ways analagous to the behavior of Galactic Black Hole
Candidates (GBHCs) in the sense that GBHCs in their high state are similar
to NLS1s while those in the low state are similar to BLS1s (e.g., Pounds
et al.\ 1995).
As it has been argued that the high-state GBHCs represent accretion near
or above the Eddington rate, this may suggest that the extreme
properties of NLS1s also result from high accretion rates (e.g., Vaughan
et al.\ 1999).
 
This analogy is good regarding the spectral similarities between Seyfert
1s and NLS1s, but it is generally not appreciated that (at least based on
our relatively poorly defined current knowledge base), it breaks down when
variability is considered.
It appears that the soft X-ray variability of NLS1s is stronger than that
of BLS1s (e.g., Turner et al.\ 1999), but for GBHCs, it is the low-state
objects that have the stronger variability.
(In fact, there is also the behavior in the very high state 
(Van Paradijs 1998), but this is not directly addressed in this analogy.)

In this study, the initial results from a systematic hard X-ray variability 
survey undertaken with the Rossi X-ray Timing Explorer (RXTE) were studied.
This work focused on BLS1s but did include one NLS1 (Akn 564), as the
steep-spectrum NLS1s are much weaker and less numerous than BLS1s in the
hard X-rays.
As discussed below, the NLS1 showed behavior that was not what was naively
expected on the basis of the soft X-ray results.
However, we emphasize that this is a very preliminary result based on a
single NLS1, so no firm conclusions can yet be drawn.
 
\section{Data}

As part of a program to study the fluctuation power density (PDS) of
Seyfert 1 galaxies, even sampling (every 4.3 d) was obtained for five BLS1s 
(NGC 3516, NGC 3783, NGC 4151, NGC 5548 and Fairall 9) and 1 NLS1 (Akn 564).
As of this writing, the first 9 months of data have been processed by the RXTE Science Operations Center and used in this contribution.

\section{Light Curves and Excess Variance Analysis}

These data were analyzed to produce 16 s light curves by standard
techniques, as discussed, e.g., in Edelson et al.\ (2000).
These data were then binned up by orbit to produce the light curves shown
in Figure 1.
The excess variance (e.g., Turner et al.\ 1999) has been computed for each
of these data sets (gathered with near-identical sampling) and listed to
the right of each panel.
Note that the excess variance of the one NLS1, Akn 564, is by no means
extreme, but is instead near the middle of the distribution of the BLS1s.
 
\begin{figure}[htb]
\centerline{\psfig{figure=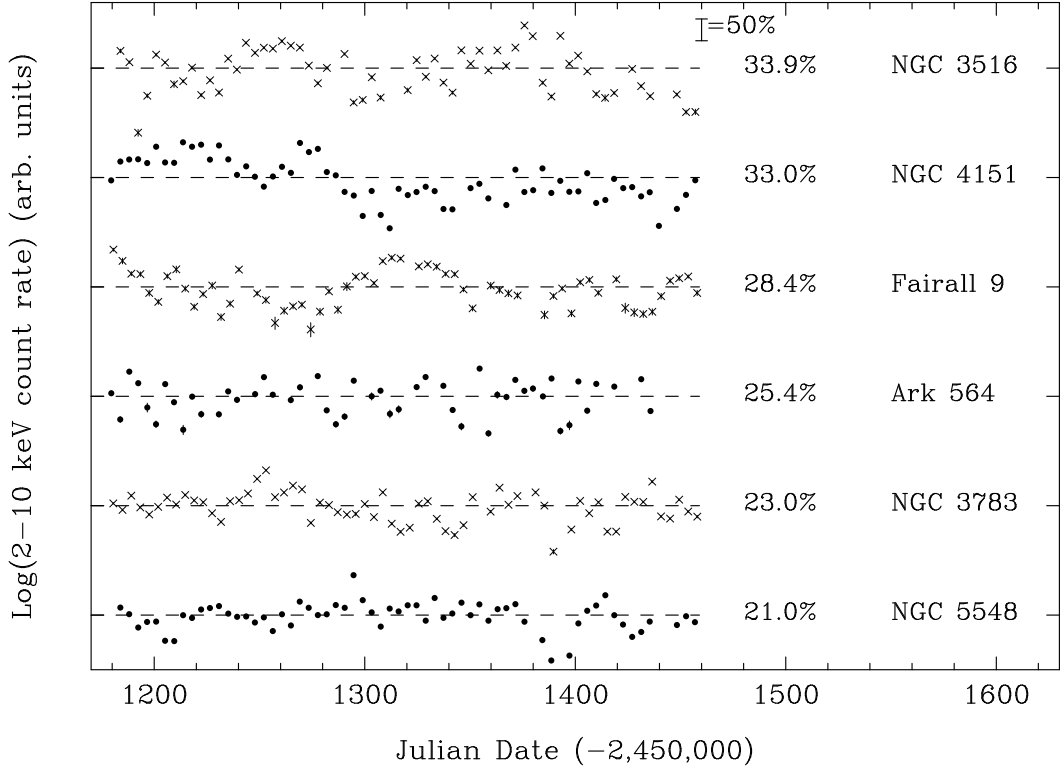,height=4.0truein,width=5.5truein,angle=270}}
\caption{Light curves for the six target Seyfert galaxies.  Note that the
one NLS1, Akn 564, does not show unusually strong hard X-ray variability.}
\end{figure}
 
One might have expected the NLS1 to have a much larger excess
variance, as indeed has been reported in the short time scale, softer
ROSAT and ASCA data (e.g., Turner et al.\ 1999).
There are a number of possible explanations of this, the most obvious
being the strong caution that this has only been seen in a single object.
In spite of the difficulties in making such a study in the hard X-rays, as
discussed below, it is expected that similarly-sampled RXTE observations
of a second NLS1, Ton S180, should be available soon.

Two additional differences with the ASCA (or ROSAT) data could have 
contributed to this as well:
First, those satellites sampled the softer X-rays, and it is in the soft
X-rays that NLS1s and BLS1s show their greatest spectral slope divergance
as well.
Second, the fact that RXTE was designed and executed with much more
relaxed and cooperative scheduling constraints makes it more well-suited
to study the longer time scales.
It is possible that processes that dominate the short time scales could be
different than those that are important over longer time scales.
 
\section{Spectral Variability}

A second unexpected but tentative result is shown in Figure 2, which plots
the relatively softer RXTE band (2-4 keV) against the harder 7-10 keV
band.
(The 4-7 keV band was avoided to minimize the contribution from any iron
line.)
All of the BLS1s show stronger variability at softer energies, while the
NLS1 shows the opposite: stronger variability in the harder bands.
This is not consistent with the idea that the soft X-rays are more
strongly variable for NLS1s, but there is a possible instrumental problem
because the NLS1 Akn 564 (about 1 c/s) is much fainter than the BLS1s .
Unfortunately, there are some serious problems with the background models
as a function of energy, which is important because RXTE is a non-imaging
instrument.
All of the BLS1s show stronger variability at softer energies, while the
NLS1 shows the opposite: stronger variability in the harder bands.
Sufficient gaps exist in the current understanding of the RXTE background 
model to make it impossible to rule out systematic errors as the cause of 
this effect.

\begin{figure}[htb]
\centerline{\psfig{figure=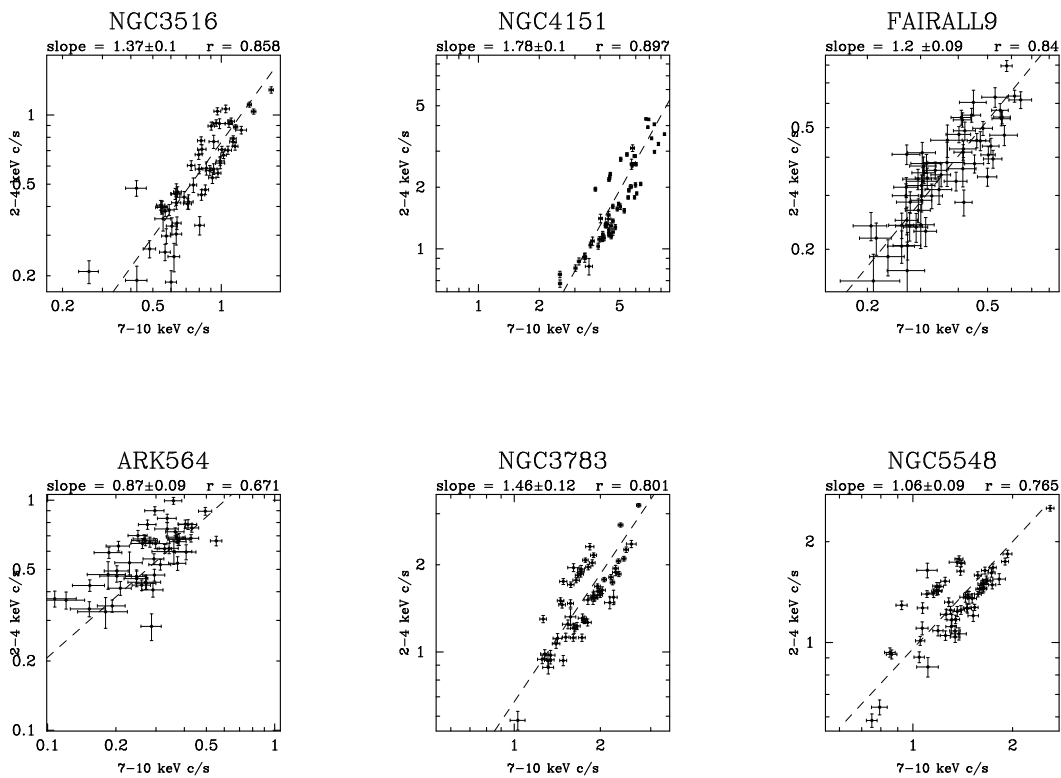,height=4.0truein,width=5.5truein,angle=270}}
\caption{Plots of hard and soft band count rates for the six target Seyfert galaxies.  Note that the one NLS1 shows stronger variability in the hardest band.}
\end{figure}


\begin{thebibliography}{999}

\bibitem{1} Boller, T., Brandt, N.\ \& Fink, H.\ 1996, A\&A, 305, 5

\bibitem{2} Edelson, R. et al.\ 2000, ApJ, in press, astro-ph/9912266

\bibitem{3} Pounds, K. et al.\ 1995, MNRAS, 277, L5

\bibitem{4} Turner, T.\ J.\ et al.\ 1999, ApJ, 510, 178   

\bibitem{5} Van Paradijs, J.\ 1998, {\it The Many Faces of Neutron Stars}, eds.\ 
R.\ Buccheri, J.\ van Paradijs, \& M.\ A.\ Alpar (Boston:Kluwer), p.\ 279

\bibitem{6} Vaughan, S.\ et al. 1999, MNRAS, 309, 113

\end{thebibliography}
\end{document}